\documentclass{article}
\usepackage{spconf,amsmath,graphicx}
\usepackage{amsfonts}
\usepackage{multirow}
\usepackage{xcolor}
\usepackage{hyperref}
\usepackage{nicefrac}
\usepackage{booktabs}
\usepackage{makecell}
\usepackage[subtle]{savetrees}
\usepackage{subcaption}
\usepackage{graphicx}
\usepackage{kantlipsum}
\usepackage[capitalize,nameinlink]{cleveref}

\newcommand{\bee}{\begin{eqnarray}}
\newcommand{\eee}{\end{eqnarray}}

\newlength{\widebarargwidth}
\newlength{\widebarargheight}
\newlength{\widebarargdepth}

\newcommand{\eat}[1]{}

\newcommand{\btx}{\tilde{\mathbf{x}}}
\newcommand{\bty}{\tilde{y}}

\newcommand{\bx}{\mathbf{x}}

\newcommand{\real}{\mathbb{R}} 

\usepackage{multirow}
\def\tablespace{\vspace{10pt}}

\newcommand{\newpara}[1]{\vspace{5pt}\noindent\textbf{#1}}

\title{Cross attentive pooling for speaker verification}
\name{Seong Min Kye$^{1,2}$, Yoohwan Kwon$^{1,3}$, Joon Son Chung$^{1}$}

\address{$^1$Naver Corporation, $^2$Korea Advanced Institute of Science and Technology, $^3$Yonsei University}
\begin{document}

\maketitle

\begin{abstract}
The goal of this paper is text-independent speaker verification where utterances come from `in the wild' videos and may contain irrelevant signal. While speaker verification is naturally a pair-wise problem, existing methods to produce the speaker embeddings are instance-wise. In this paper, we propose \emph{Cross Attentive Pooling} (CAP) that utilises the context information across the reference-query pair to generate utterance-level embeddings that contain the most discriminative information for the pair-wise matching problem. Experiments are performed on the VoxCeleb dataset in which our method outperforms comparable pooling strategies.
\end{abstract}

\vspace{10pt}
\noindent\textbf{Index Terms}: speaker recognition, speaker verification, cross attention.

\section{Introduction}
\label{sec:intro}

Automatic speaker recognition is an attractive way to verify someone’s identity since the voice of a person is one of the most easily accessible biometric information. Due to this non-invasive nature and the technological progress, speaker recognition has recently gained considerable attention both in the industry and in research. 

While the definition of speaker recognition encompasses both identification and verification, the latter has more practical applications -- for example, the use of speaker verification is becoming popular in call centres and in AI speakers. Unlike closed-set identification, open-set verification aims to verify the identity of speakers unseen during training. Therefore, speaker verification is naturally a metric learning problem in which voices must be mapped to representations in a discriminative embedding space. 

While mainstream literature in the field have learnt speaker embeddings via the classification loss~\cite{Nagrani17,snyder2017deep,snyder2018x, kwon2020intra,heigold2016end}, such objective functions are not designed to optimise embedding similarity. More recent works~\cite{jung2020improving,Xie19a,hajibabaei2018unified,liu2019large,garcia2019x,zeinali2019but,xiang2019margin} have used additive margin variants of the softmax function~\cite{wang2018additive,wang2018cosface,deng2019arcface} to enforce inter-class separation which has been shown to improve verification performance.

Since open-set verification addresses identities unseen during training, it can be formulated as a few-shot learning problem where the network should recognise unseen classes with limited examples. Prototypical networks~\cite{snell2017prototypical} have been proposed in which the training mimics the few-shot learning scenario, and this strategy has recently shown to achieve competitive performance in speaker verification~\cite{kye2020meta,wang2019centroid,anand2019few,chung2020defence,huh2020augmentation}. 

In order to train networks to optimise the similarity metric, frame-level features (or representations) produced must first be aggregated into an utterance-level feature. A na\"ive way to produce an utterance-level embedding is to take a uniformly weighted average of the frame-level representations, which is referred to as Temporal Average Pooling (TAP) in the existing literature. Self-Attentive Pooling (SAP)~\cite{cai2018exploring} has been proposed to pay more attention to the frames that are more discriminative for verification. However, the instance-level self-attention finds the features that are more discriminative for speaker verification in general (i.e. across the whole training set) rather than for the specific examples in the support set.

In few-shot learning, cross attention networks (CAN)~\cite{hou2019cross} have been recently proposed to select attention based on unseen target classes, by attending to the parts of the input image that is relevant and discriminative to the examples in the support set. This idea is applicable to speaker verification, since the features that are discriminative for comparing an utterance against one class (or speaker) in the support set may be different to the features for comparing against another class. 

To this end, we propose \emph{Cross Attentive Pooling} (CAP) which computes the attention with reference to the example in the support set in order to effectively aggregate frame-level information into an utterance-level feature. In this way, the network is able to identify and focus on the parts of the utterance that provide characterising features for the particular class in the support set. 
This is similar to how humans tend to look for common characterising features between the pair of samples when recognising instances from unseen classes, whether these are speakers or visual objects.
Unlike instance-level pooling, the proposed attention module takes full advantage of the pair-wise nature of the verification task, by modelling the relevance between the class (prototype) feature  and the query feature.

The effectiveness of our method is demonstrated on the popular VoxCeleb dataset \cite{Nagrani19} in which we report improvements over existing pooling methods.

\section{Methods}

\subsection{Few-shot learning framework}

We use a few-shot learning framework in order to train the embeddings for speaker recognition. In particular, our implementation is based on the prototypical networks~\cite{snell2017prototypical}, which have been shown to perform well in speaker verification~\cite{kye2020meta,chung2020defence}.

 \newpara{Batch formation.}
 Each mini-batch contains a support set $\mathcal{S}$ and a query set $\mathcal{Q}$. A mini-batch contains $M$ utterances from each of $N$ different speakers. We use a single utterance for each speaker in the support set $\mathcal{S} = \{(\bx_i,y_i)\}_{i=1}^{N \times 1}$ and the rest of the utterances ($2 \leq i \leq M$) in the query set  $\mathcal{Q} = \{(\btx_i,\bty_i)\}_{i=1}^{N \times (M-1)}$, where $y,\bty \in \{1,\dots,N\}$ is the class label.
 
 \newpara{Training objective.}
Since the support set is formed from a single utterance $\bx$, the prototype (or centroid) is the same as the support utterance for each speaker $y$:
\begin{equation}
P_{y} = \bx
\label{eqn:proto_cent}
\end{equation}

\noindent The cross-entropy loss with a log-softmax function is used to minimise the distance between segments from same speaker and maximise the distance between different speakers. 

\begin{equation}
L_{NP} = -\frac{1}{|\mathcal{Q}|} \sum_{(\btx,\bty) \in \mathcal{Q}} \log
\frac{e^{d(\btx, P_{\bty})}}
{\sum_{y=1}^N e^{d(\btx, P_{y})}}
\label{eqn:proto_loss}
\end{equation}
 
\noindent We use the same distance metric as \cite{kye2020meta}, where the distance function is the cosine similarity between the prototype and the query with the scale of the query embedding.

\begin{align}
    d(\btx, P_{\bty}) = \frac{\btx^T P_{\bty}}{\|P_{\bty}\|_2} =\|\btx\|_2 \cdot cos(\btx, P_{\bty})
    \label{eq:dist_metric}
\end{align}

\noindent We refer to the prototypical loss with this similarity function as the {\bf Normalised Prototypical (NP)} loss in the rest of this paper.

In \cite{kye2020meta}, they use prototypical loss with softmax loss which they call \emph{global classification}. The softmax loss classifies samples in each mini-batch for all classes of training set.
\begin{equation}
L_s = -\frac{1}{B}\sum_{i=1}^{B} log \frac{e^{d(\bx_i,w_{y_i})}}{\sum_{c=1}^C e^{d(\bx_i,w_c)}}
\end{equation}
The batch size is $B=|\mathcal{S}|+|\mathcal{Q}|$ and $w$ is the set of weights for all classes. In addition, the distance metric is used the same as Eq.~\ref{eq:dist_metric}.  The final objective is the sum of NP and the softmax loss with equal weighting, as follows:
\begin{align}
L = L_{NP} + L_s
\end{align}
By incorporating NP and the softmax loss, we can train the embeddings to be discriminative over all classes, as opposed to only the classes in the mini-batch.

\subsection{Instance-wise aggregation}
\label{sec:agg}

An ideal utterance-level feature should be invariant to temporal position, but {\em not} frequency. Since 2D convolutional neural networks~\cite{Chatfield14,He16} produce 2D activation maps, \cite{Nagrani17} has proposed aggregation layers that are fully connected only along the frequency axis. This produces a $1 \times T$ feature map before the pooling layers, which are described in the following sections.

\newpara{Temporal Average Pooling (TAP).} 
The TAP layer simply takes the mean of the features along the time domain. 

\begin{equation}
e = \frac{1}{T} \sum_{n=1}^{T} x_t
\end{equation}

\newpara{Self-Attentive Pooling (SAP).}
In contrast to the TAP layer that pools the features over time with uniform weights, the self-attentive pooling (SAP) layer~\cite{cai2018exploring,bhattacharya2017deep,rahman2018attention} pays attention to the frames that are more informative for frame-level speaker recognition. 

The frame-level features $\{x_1, x_2, \dots, x_T\}$ are first mapped to hidden features $\{h_1, h_2, \dots, h_T\}$ using a single layer perceptron with learnable weights $W$ and $b$.

\begin{equation}
h_t = tanh(Wx_t + b)
\label{eq:w1}
\end{equation}

\noindent The similarity between the hidden features and a learnable context vector $\mu$ is computed, which represents the relative importance of the hidden features. The context vector can be seen as a high-level representation of what makes the frames informative for speaker recognition.

\begin{equation}
w_t = \frac{exp(h_t^T \mu)}{\sum_{t=1}^T exp(h_t^T \mu)}
\label{eq:w2}
\end{equation}

\noindent The utterance-level feature $e$ can be obtained as a weighted sum of the frame-level representations.

\begin{equation}
e = \sum_{t=1}^{T} w_t x_t
\label{eq:weightedsum}
\end{equation}

\begin{figure*}[t]
\centering
\includegraphics[width=0.9\linewidth]{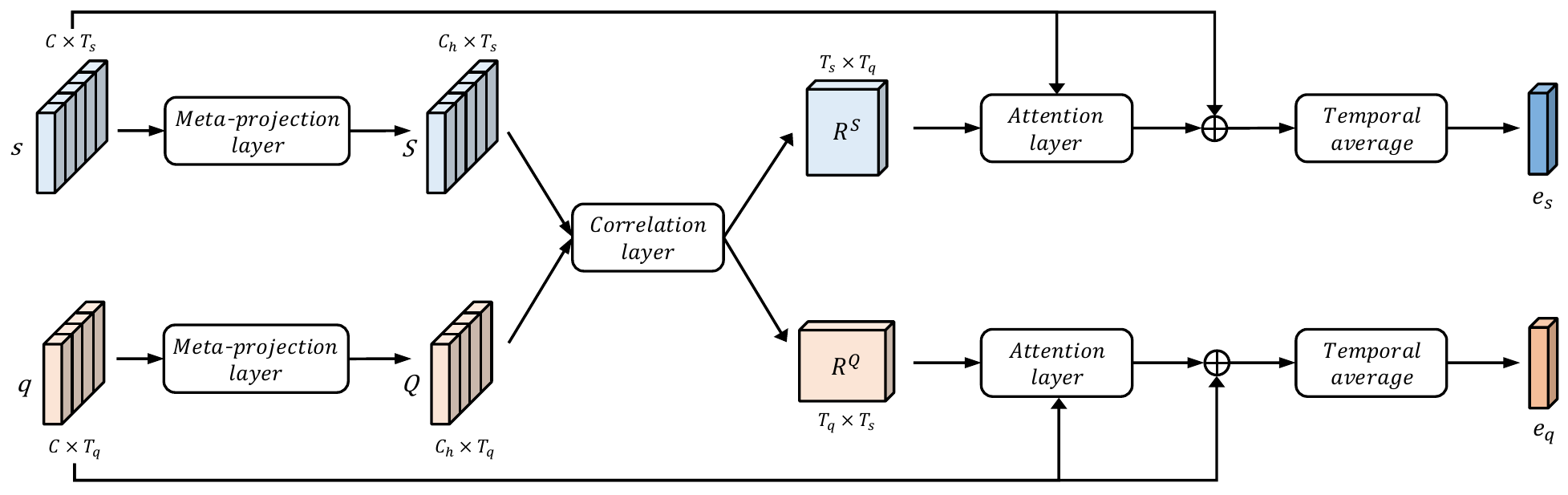}
\caption{The procedure of \emph{Cross Attentive Pooling} (CAP).}
\label{fig:overview}
\end{figure*}

\begin{figure}[t]
\centering
\includegraphics[width=0.9\linewidth]{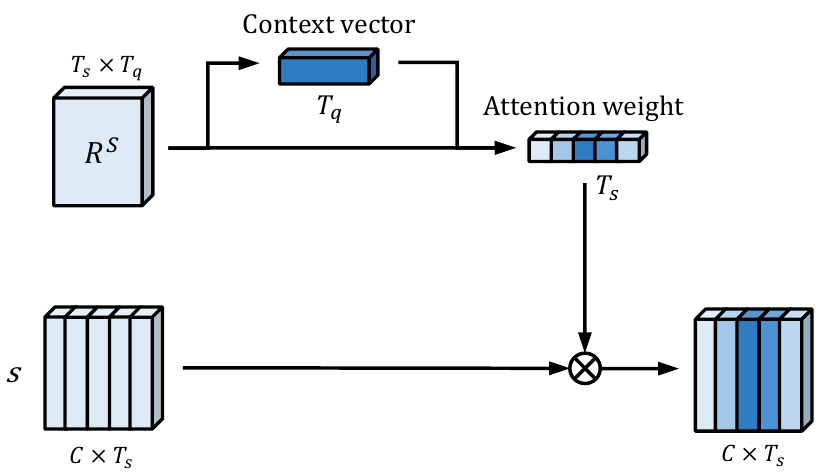}
\caption{Attention layer.}
\label{fig:attention}
\end{figure}

\subsection{Pair-wise aggregation}
Unlike traditional instance-wise aggregation, our proposed method aggregates frame-level features, utilising information of the frame features from the other utterance. In order to match the objective in training and testing, we use the prototypical networks~\cite{snell2017prototypical}, which is metric-based meta-learning framework. In this framework, we train our cross attentive pooling (CAP) using the pairs of support and query set. In the test scenario, the support and the query sets correspond to the enrollment and the test utterances, respectively.

For every pair of utterances from the query and the support sets, we extract frame-level features $s = \{s_1, s_2, \dots, s_{T_{s}}\}$ and $q = \{q_1, q_2, \dots, q_{T_{q}}\}$. Then, with the meta-projection layer $g_{\phi}(\cdot)$, we  extract hidden features from the frame-level features. This non-linear projection allows us to quickly adapt to arbitrary frames, so that the similarity of the frame pair can be well measured. The layer consists a single-layer perceptron followed by a ReLU activation function.
\begin{equation}
g_{\phi}(\cdot) = \text{max}\big(0, W(\cdot)+b\big)
\end{equation}
After the meta-projection layer, we can obtain $S = \{S_i\}_{i=1}^{T_s}$ and $Q = \{Q_i\}_{i=1}^{T_q}$ as the hidden features for every frame, where $S_i$ and $Q_i$ denotes $g_{\phi}(s_i)$ and $g_{\phi}(q_i)$, respectively.

\newpara{Correlation matrix.}
Correlation matrix $R$ summarises similarity for every possible pair of frames. 
$R^S \in \real^{T_s \times T_q}$ is computed as:
\begin{equation}
R^S_{i,j} = \bigg(\frac{S_i}{\|S_i\|_2}\bigg)^{T}\bigg(\frac{Q_j}{\|Q_j\|_2}\bigg)
\end{equation}
Note that $R^Q = (R^S)^T$. 

\newpara{Pair-adaptive attention.}
In order to obtain the pair-adaptive context vector, we average correlation matrix along with its own time axis as:
\begin{equation}
\mu_s=\frac{1}{T_s}\sum_{i=1}^{T_s} R^S_{i, *}
\end{equation}
where $\mu_s \in \real^{T_s}$ and $R^S_{i,*}$ denotes $i$-th row vector. Each row vector has the information of similarity to all frames of the other utterance. Therefore, the average correlation for each frame of the other utterance can be presented by $\mu$, which is used in the context vector to determine how similar it is to the other utterance.

The attention weights are given by the following equation for every utterance.
\begin{equation}
w^{s}_{t}=\frac{exp((\mu_s^T R^S_{t,*})/\tau)}{\sum_{i=1}^{T_s}exp((\mu_s^T R^S_{i,*})/\tau)}
\end{equation}
where $\tau$ is temperature scaling, which sharpens attention distribution.
\begin{equation}
e_s = \frac{1}{T_s}\sum_{t=1}^{T_s}(1+w^{s}_{t}) s_t
\end{equation}
As done in \cite{hou2019cross}, we use a residual attention mechanism to obtain the utterance-level feature. For the other utterance, $e_q$ can be obtained in the same way.

After the aggregation, we finally obtain the speaker embedding as follows:
\begin{equation}
\bx = We
\end{equation}
where $W$ is projection matrix for embedding space.

\newpara{Global classification for CAP.} Cross attentive pooling cannot be applied directly to the standard softmax-based training, since there is no ‘pair' in the classification task. With the CAP, we apply global classification only for the positive (same class) pairs. For instance, $M-1$ embeddings are generated by query utterance of the same class for each support utterance, whereas a single embedding is generated by support utterance for each query utterance. To summarise, the global classification is applied to $M-1$ support embeddings and $M-1$ query embeddings for each class ({\em i.e.} $2N(M-1)$ in total). It reduces the variance of support embeddings made by the utterance pairs of the same class, so that attention can be more focused on the frames with abundant speaker information.
\begin{table}[t]
\renewcommand\arraystretch{1.1}
\centering
\caption{Fast ResNet-34 architecture. ReLU and batchnorm layers are not shown.  Each row specifies the number of convolutional filters, their sizes and strides as {\bf size $\times$ size, \# filters, stride}. The output from the fully connected layer is ingested by the pooling layers.}
\vspace{-0.15in}
\label{table:convnet}
\tablespace
\resizebox{0.48\textwidth}{!}{
	\small
\begin{tabular}{ c|c|c }
\hline
\textbf{layer name}   & \textbf{Filters} & \textbf{Output}  \\ \hline
conv1 & \makecell{$7 \times 7,16$, stride 2 \\ $ 3 \times 3$, Maxpool, stride 2}  & \makecell{$20 \times T \times 16$} \\   \hline
conv2  & $\begin{bmatrix}  3 \times 3, 16 \\ 3 \times 3, 16 \end{bmatrix} \times 3 $, stride 1 & $20 \times T \times 16$        \\\hline
conv3 & $\begin{bmatrix}  3 \times 3,32 \\ 3 \times 3,32 \end{bmatrix} \times 4 $, stride 2 & $10 \times \nicefrac{T}{2} \times 32$    \\\hline
conv4 & $\begin{bmatrix}  3 \times 3,64 \\ 3 \times 3,64 \end{bmatrix} \times 6 $, stride 2 & $5 \times \nicefrac{T}{4} \times 64$      \\\hline
conv5  & $\begin{bmatrix}  3 \times 3,128 \\ 3 \times 3,128 \end{bmatrix} \times 3 $, stride 2 & $5 \times \nicefrac{T}{4} \times 128$      \\\hline
pool &  $ 9 \times 1$ &     $1 \times \nicefrac{T}{4} \times 128$ \\\hline
aggregation & TAP {\em or} SAP {\em or} CAP & $1 \times 128$\\\hline
fc & FCN, 512 & $1 \times 512$\\\hline
\end{tabular}
}
\end{table}
\section{Experiments}
\subsection{Input representations}
During training, we randomly extract 2-second fixed-length temporal segments from each utterance.
Spectrograms are extracted with a hamming window of width 25ms and step 10ms. 40-dimensional Mel filterbanks are used as the input to the network.
Mean and variance normalisation (MVN) is performed with instance normalisation~\cite{ulyanov2016instance}. For each mini-batch, we set the number of speaker $N$ and utterance for each speaker $M$ to 200 and 3, respectively. Since the VoxCeleb dataset contains continuous speech, voice activity detection (VAD) is not used during training and testing. Aside from taking the 2-second randomsegments, no data augmentation is performed during training or testing.

\subsection{Trunk architecture}

Experiments are performed using the Fast ResNet-34 architecture introduced in~\cite{chung2020defence}. Residual networks~\cite{He16} are used widely in image recognition and has recently been applied to speaker recognition~\cite{Xie19a,cai2018exploring,Chung18a}. Fast ResNet-34 is the same as the original ResNet with 34 layers, except with only one-quarter of the channels in each residual block in order to reduce computational cost (i.e. 16-32-64-128 channels for each layer). The model only has 1.4 million parameters compared to 22 million of the standard ResNet-34, and minimises the computation cost by reducing the activation maps early in the network. The network architecture is given in Table~\ref{table:convnet}. 

For the meta-projection layer, we use a single fully-connected layer with  128 dimensions, which is followed by ReLU activation function. We use the temperature scaling $\tau$ with a fixed value of 0.05.

\begin{table*}[t]
\renewcommand\arraystretch{1.1}
\centering
    \caption{Comparison with previous the state-of-the-art works. {\bf NP}: Normalised Prototypical, {\bf AP}: Angular Prototypical, {\bf TAP}: Temporal average pooling, {\bf SAP}: Self-attentive pooling, {\bf CAP}: Cross attentive pooling.
    The experiments are repeated three times and the mean and the standard deviation are reported.}
\resizebox{1.\linewidth}{!}{
 \begin{tabular}{c c c c | c c | c c }
\toprule 
\multirow{2}{*}{Feature extractor} & \multirow{2}{*}{Feature} & \multirow{2}{*}{Objective} & \multirow{2}{*}{Aggregation} & \multicolumn{2}{c}{VoxCeleb} &
\multicolumn{2}{c}{VOiCES}\\
 &  &  &  & 
MinDCF & EER (\%) & MinDCF & EER (\%)\\ \midrule
Thin ResNet-34~\cite{Xie19a} & Spec-257 & Softmax & GhostVLAD &  - & 3.22 & - & - \\
ResNet-50~\cite{yu2019ensemble} & Spec-512 & EAM-Softmax & TAP & - & 2.94 & - & - \\
ResNet-34~\cite{SPE} & MFB-64 & A-Softmax & SPE &  - & 2.61 & - & - \\
Fast ResNet-34~\cite{chung2020defence} & MFB-40 & AP & TAP &  0.176 & 2.22 & - & - \\ 
ResNet-34~\cite{kye2020meta} & MFB-40 & NP + Softmax & TAP &  - & 2.08 & - & - \\ \midrule

Fast ResNet-34 & MFB-40 & NP + Softmax & TAP & 0.154\scriptsize$\pm$0.054 &
2.08\scriptsize$\pm$0.01 &
0.301\scriptsize$\pm$0.005 &
4.04\scriptsize$\pm$0.17 \\
Fast ResNet-34 & MFB-40 & NP + Softmax & SAP &  0.156\scriptsize$\pm$0.071 &
2.09\scriptsize$\pm$0.04 &
0.315\scriptsize$\pm$0.011 &
4.04\scriptsize$\pm$0.20 \\ 
Fast ResNet-34 & MFB-40 & NP + Softmax & CAP & \textbf{0.148\scriptsize$\pm$0.010} &
\textbf{1.88\scriptsize$\pm$0.07} &
\textbf{0.291\scriptsize$\pm$0.024} &
\textbf{3.86\scriptsize$\pm$0.19} \\ \bottomrule 

Fast ResNet-34 & MFB-64 & NP + Softmax & TAP & 0.146\scriptsize$\pm$0.007 &
1.95\scriptsize$\pm$0.02 &
0.270\scriptsize$\pm$0.004 &
3.65\scriptsize$\pm$0.13 \\
Fast ResNet-34 & MFB-64 & NP + Softmax & SAP & 0.137\scriptsize$\pm$0.010 &
1.91\scriptsize$\pm$0.04 &
0.256\scriptsize$\pm$0.012 &
3.50\scriptsize$\pm$0.09 \\ 
Fast ResNet-34 & MFB-64 & NP + Softmax & CAP & \textbf{0.122\scriptsize$\pm$0.007} &
\textbf{1.74\scriptsize$\pm$0.08} &
\textbf{0.247\scriptsize$\pm$0.005} &
\textbf{3.39\scriptsize$\pm$0.12} \\ \bottomrule 

 \end{tabular}
}
    \label{tbl:full}
\end{table*}

\begin{table}[ht]
\centering
    \caption{Ablation study on CAP. Temp: Temperature; MPL: Meta-projection layer; GC: Global classification.}
 \begin{tabular}{c c c| c c }
 \hline
 Temp & MPL & GC & MinDCF & EER\% \\
 \hline
 \hline
 &  &  &
0.251\scriptsize$\pm$0.011 & 3.22\scriptsize$\pm$0.01 \\\hline
\checkmark &  &  &
0.235\scriptsize$\pm$0.010 & 3.02\scriptsize$\pm$0.04\\\hline
\checkmark & \checkmark &  &
0.236\scriptsize$\pm$0.034 & 2.84\scriptsize$\pm$0.01 \\\hline
\checkmark & \checkmark & \checkmark &
\textbf{0.148\scriptsize$\pm$0.010} & \textbf{1.88\scriptsize$\pm$0.07} \\ \hline
 \hline
 \end{tabular}
    \label{tbl:ablation}
\end{table}

\begin{figure*}[t]
\centering
\includegraphics[width=1.0\linewidth]{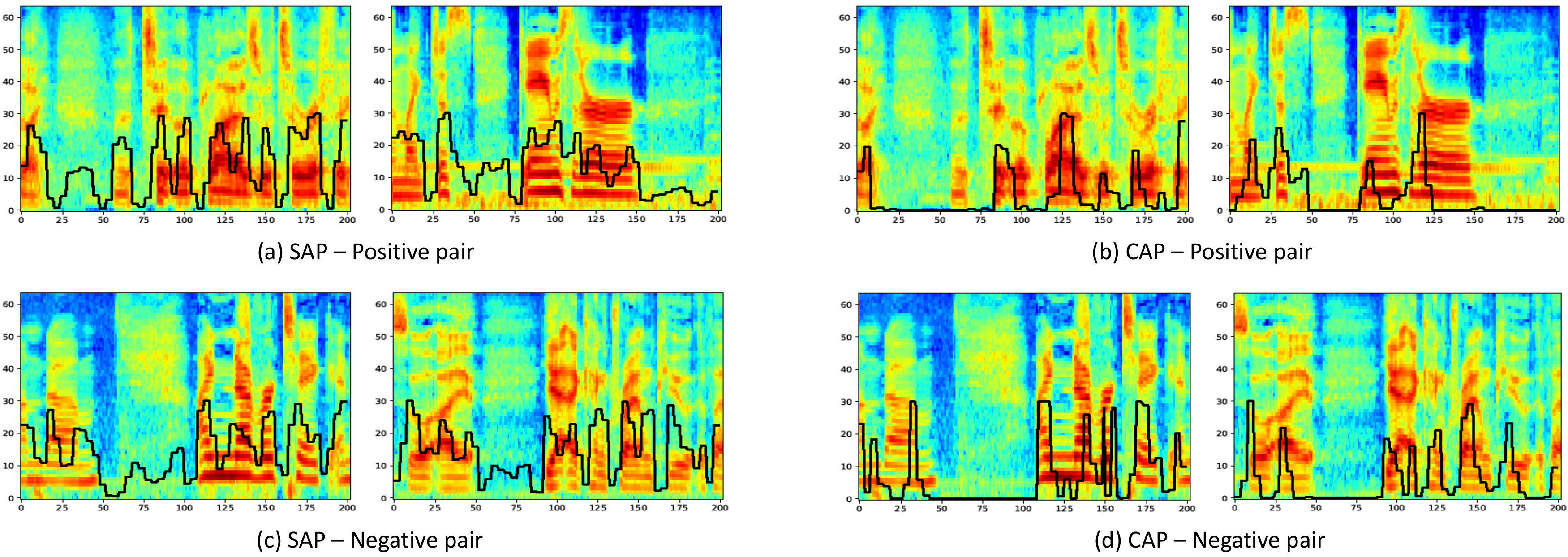}
\caption{Mel filterbank overlaid with attention distribution.}
\label{fig:visualization}
\end{figure*}

\subsection{Implementation details}
\label{subsec:impl}

\newpara{Datasets.}
The networks are trained on the development set of VoxCeleb2~\cite{Chung18a} and tested on the original test set of VoxCeleb1~\cite{Nagrani17}. Both are for large-scale text-independent speaker recognition datasets, containing 5,994 and 1,251 speakers respectively. Note that there is no overlap between the development set of
VoxCeleb2 dataset and the VoxCeleb1 dataset. To further verify effectiveness on out-of-domain data, we use the VOiCES dataset~\cite{richey2018voices} which contains 4 million pairs from 15.904 utterances.

\newpara{Training.}
Our implementation is based on the PyTorch framework~\cite{paszke2019pytorch}.
The models are trained using a NVIDIA V100 GPU with 32GB memory for $500$ epochs. 
The networks are trained with the SGD optimizer with the Nesterov momentum of $0.9$, and weight decay is set to $0.0001$. We use an initial learning rate of $0.1$ and decay it by a factor of 10 until convergence with patience of 10. The networks take 2 days to train using a single GPU.


\subsection{Evaluation}
\label{subsec:results}
\newpara{Evaluation protocol.}
We report two performance metrics: (1) the Equal Error Rate (EER) which is the rate at which both acceptance and rejection errors are equal; and (2) the minimum of the detection cost function function used by the NIST SRE~\cite{nist2018}
and the VoxCeleb Speaker Recognition Challenge (VoxSRC)~\cite{chung2019voxsrc} evaluations. 
In order to compute the EER, we sample 10 3.5-second speech segments at regular time intervals from each utterance and compute the mean of $10 \times 10 = 100$ distances from all possible combinations per each pair.
This protocol is in line with that used by~\cite{huh2020augmentation, Chung18a}.
The parameters $C_{miss}=1$, $C_{fa}=1$ and $P_{target}=0.05$ are used for the cost function, same as that used in the VoxSRC.

In order to mitigate the effect of random initialisation, we repeat all experiments three times and report mean and standard deviation of the results.

\newpara{Comparison with recent works.}
In Table~\ref{tbl:full}, we compare our method with other existing state-of-the-art models. As shown in \cite{kye2020meta,chung2020defence}, we see that prototypical-based models outperform other baselines. It should be noted that we use a model that has smaller number of parameter (1.4 million). For the VoxCeleb dataset, our method outperforms TAP and SAP. Especially, CAP outperforms SAP with $10.05\%$ using comparable method and architecture. The baseline models use instance-wise aggregation method, where a single utterance is aggregated without interaction with other utterances. In contrast, our method uses frame-level information of other utterances. It allows to quickly adapt to unseen utterance using rich information rather than instance-wise aggregation methods. Our model also consistently ourperforms baseline models on the VOiCES dataset.

\newpara{Ablation study.}
In Table~\ref{tbl:ablation}, we show the results of an ablation study on the VoxCeleb2 dataset. We see that the temperature has a very important role, allowing the attention to focus more on the informative frames. Next, the meta-projection layer also contributes to performance improvements, by helping the CAP to be trained with learnable parameters and quickly adapt to frame-level features of unseen utterances. Lastly, we observe that the global classification leads a large performance improvement. Since it is applied only to the embeddings generated by utterance of the same class, it allows intra-class variance can be effectively reduced.

\newpara{Qualitative analysis.}
Figure~\ref{fig:visualization} visualises the effect of attention on the Mel filterbank, and compares CAP to SAP. For CAP, we show positive and negative pairs. In both cases, attention is well activated in the speech section. However, we observe that SAP gets high activation even in areas with weak speech signals. In contrast, the proposed CAP clearly reduces activation weight on the non-speech periods. Moreover, attention is effectively assigned to a negative pair (see Figure~\ref{fig:visualization}d). The better attention can help explain why CAP has superior performance in speaker verification.

\section{Conclusion}
\label{sec:conc}
In this paper, we propose novel pair-wise aggregation method for speaker verification, cross attentive pooling. In contrast to existing instance-based methods, the pair-wise strategy benefits from the contextual information by looking at the parts of the speech pair. 
The pair-wise pooling method is not only applicable to the prototypical framework, but also to other metric learning objectives such as the contrastive loss. 


\clearpage
\bibliographystyle{IEEEbib}
\bibliography{shortstrings,vgg_local,vgg_other,refs}

\begin{thebibliography}{10}

\bibitem{Nagrani17}
Arsha Nagrani, Joon~Son Chung, and Andrew Zisserman,
\newblock ``{VoxCeleb}: a large-scale speaker identification dataset,''
\newblock in {\em Proc. Interspeech}, 2017.

\bibitem{snyder2017deep}
David Snyder, Daniel Garcia-Romero, Daniel Povey, and Sanjeev Khudanpur,
\newblock ``Deep neural network embeddings for text-independent speaker
  verification.,''
\newblock in {\em Proc. Interspeech}, 2017, pp. 999--1003.

\bibitem{snyder2018x}
David Snyder, Daniel Garcia-Romero, Gregory Sell, Daniel Povey, and Sanjeev
  Khudanpur,
\newblock ``X-vectors: Robust dnn embeddings for speaker recognition,''
\newblock in {\em Proc. ICASSP}. IEEE, 2018, pp. 5329--5333.

\bibitem{kwon2020intra}
Yoohwan Kwon, Soo-Whan Chung, and Hong-Goo Kang,
\newblock ``Intra-class variation reduction of speaker representation in
  disentanglement framework,''
\newblock in {\em Proc. Interspeech}, 2020.

\bibitem{heigold2016end}
Georg Heigold, Ignacio Moreno, Samy Bengio, and Noam Shazeer,
\newblock ``End-to-end text-dependent speaker verification,''
\newblock in {\em Proc. ICASSP}. IEEE, 2016, pp. 5115--5119.

\bibitem{jung2020improving}
Youngmoon Jung, Seong~Min Kye, Yeunju Choi, Myunghun Jung, and Hoirin Kim,
\newblock ``Improving multi-scale aggregation using feature pyramid module for
  robust speaker verification of variable-duration utterances,''
\newblock in {\em Interspeech}, 2020.

\bibitem{Xie19a}
Weidi Xie, Arsha Nagrani, Joon~Son Chung, and Andrew Zisserman,
\newblock ``Utterance-level aggregation for speaker recognition in the wild,''
\newblock in {\em Proc. ICASSP}, 2019.

\bibitem{hajibabaei2018unified}
Mahdi Hajibabaei and Dengxin Dai,
\newblock ``Unified hypersphere embedding for speaker recognition,''
\newblock {\em arXiv preprint arXiv:1807.08312}, 2018.

\bibitem{liu2019large}
Yi~Liu, Liang He, and Jia Liu,
\newblock ``Large margin softmax loss for speaker verification,''
\newblock in {\em INTERSPEECH}, 2019.

\bibitem{garcia2019x}
Daniel Garcia-Romero, David Snyder, Gregory Sell, Alan McCree, Daniel Povey,
  and Sanjeev Khudanpur,
\newblock ``X-vector dnn refinement with full-length recordings for speaker
  recognition,''
\newblock in {\em Interspeech}, 2019, pp. 1493--1496.

\bibitem{zeinali2019but}
Hossein Zeinali, Shuai Wang, Anna Silnova, Pavel Mat{\v{e}}jka, and
  Old{\v{r}}ich Plchot,
\newblock ``{BUT} system description to {VoxCeleb Speaker Recognition
  Challenge} 2019,''
\newblock {\em arXiv preprint arXiv:1910.12592}, 2019.

\bibitem{xiang2019margin}
Xu~Xiang, Shuai Wang, Houjun Huang, Yanmin Qian, and Kai Yu,
\newblock ``Margin matters: Towards more discriminative deep neural network
  embeddings for speaker recognition,''
\newblock in {\em Asia-Pacific Signal and Information Processing Association
  Annual Summit and Conference}, 2019.

\bibitem{wang2018additive}
Feng Wang, Jian Cheng, Weiyang Liu, and Haijun Liu,
\newblock ``Additive margin softmax for face verification,''
\newblock {\em IEEE Signal Processing Letters}, vol. 25, no. 7, pp. 926--930,
  2018.

\bibitem{wang2018cosface}
Hao Wang, Yitong Wang, Zheng Zhou, Xing Ji, Dihong Gong, Jingchao Zhou, Zhifeng
  Li, and Wei Liu,
\newblock ``Cosface: Large margin cosine loss for deep face recognition,''
\newblock in {\em Proc. CVPR}, 2018, pp. 5265--5274.

\bibitem{deng2019arcface}
Jiankang Deng, Jia Guo, Niannan Xue, and Stefanos Zafeiriou,
\newblock ``Arcface: Additive angular margin loss for deep face recognition,''
\newblock in {\em Proc. CVPR}, 2019, pp. 4690--4699.

\bibitem{snell2017prototypical}
Jake Snell, Kevin Swersky, and Richard Zemel,
\newblock ``Prototypical networks for few-shot learning,''
\newblock in {\em NeurIPS}, 2017, pp. 4077--4087.

\bibitem{kye2020meta}
Seong~Min Kye, Youngmoon Jung, Hae~Beom Lee, Sung~Ju Hwang, and Hoirin Kim,
\newblock ``Meta-learning for short utterance speaker recognition with
  imbalance length pairs,''
\newblock in {\em Interspeech}, 2020.

\bibitem{wang2019centroid}
Jixuan Wang, Kuan-Chieh Wang, Marc~T Law, Frank Rudzicz, and Michael Brudno,
\newblock ``Centroid-based deep metric learning for speaker recognition,''
\newblock in {\em Proc. ICASSP}. IEEE, 2019, pp. 3652--3656.

\bibitem{anand2019few}
Prashant Anand, Ajeet~Kumar Singh, Siddharth Srivastava, and Brejesh Lall,
\newblock ``Few shot speaker recognition using deep neural networks,''
\newblock {\em arXiv preprint arXiv:1904.08775}, 2019.

\bibitem{chung2020defence}
Joon~Son Chung, Jaesung Huh, Seongkyu Mun, Minjae Lee, Hee~Soo Heo, Soyeon
  Choe, Chiheon Ham, Sunghwan Jung, Bong-Jin Lee, and Icksang Han,
\newblock ``In defence of metric learning for speaker recognition,''
\newblock in {\em Proc. Interspeech}, 2020.

\bibitem{huh2020augmentation}
Jaesung Huh, Hee~Soo Heo, Jingu Kang, Shinji Watanabe, and Joon~Son Chung,
\newblock ``Augmentation adversarial training for unsupervised speaker
  recognition,''
\newblock {\em arXiv preprint arXiv:2007.12085}, 2020.

\bibitem{cai2018exploring}
Weicheng Cai, Jinkun Chen, and Ming Li,
\newblock ``Exploring the encoding layer and loss function in end-to-end
  speaker and language recognition system,''
\newblock in {\em Speaker Odyssey}, 2018.

\bibitem{hou2019cross}
Ruibing Hou, Hong Chang, MA~Bingpeng, Shiguang Shan, and Xilin Chen,
\newblock ``Cross attention network for few-shot classification,''
\newblock in {\em NeurIPS}, 2019, pp. 4003--4014.

\bibitem{Nagrani19}
Arsha Nagrani, Joon~Son Chung, Weidi Xie, and Andrew Zisserman,
\newblock ``Voxceleb: Large-scale speaker verification in the wild,''
\newblock {\em Computer Speech and Language}, 2019.

\bibitem{Chatfield14}
Ken Chatfield, Karen Simonyan, Andrea Vedaldi, and Andrew Zisserman,
\newblock ``Return of the devil in the details: Delving deep into convolutional
  nets,''
\newblock in {\em Proc. BMVC}, 2014.

\bibitem{He16}
Kaiming He, Xiangyu Zhang, Shaoqing Ren, and Jian Sun,
\newblock ``Deep residual learning for image recognition,''
\newblock in {\em Proc. CVPR}, 2016.

\bibitem{bhattacharya2017deep}
Gautam Bhattacharya, Md~Jahangir Alam, and Patrick Kenny,
\newblock ``Deep speaker embeddings for short-duration speaker verification.,''
\newblock in {\em Proc. Interspeech}, 2017, pp. 1517--1521.

\bibitem{rahman2018attention}
FA~Rezaur rahman Chowdhury, Quan Wang, Ignacio~Lopez Moreno, and Li~Wan,
\newblock ``Attention-based models for text-dependent speaker verification,''
\newblock in {\em Proc. ICASSP}. IEEE, 2018, pp. 5359--5363.

\bibitem{ulyanov2016instance}
Dmitry Ulyanov, Andrea Vedaldi, and Victor Lempitsky,
\newblock ``Instance normalization: The missing ingredient for fast
  stylization,''
\newblock {\em arXiv preprint arXiv:1607.08022}, 2016.

\bibitem{Chung18a}
Joon~Son Chung, Arsha Nagrani, and Andrew Zisserman,
\newblock ``{VoxCeleb2}: Deep speaker recognition,''
\newblock in {\em Proc. Interspeech}, 2018.

\bibitem{yu2019ensemble}
Ya-Qi Yu, Lei Fan, and Wu-Jun Li,
\newblock ``Ensemble additive margin softmax for speaker verification,''
\newblock in {\em ICASSP 2019-2019 IEEE International Conference on Acoustics,
  Speech and Signal Processing (ICASSP)}. IEEE, 2019, pp. 6046--6050.

\bibitem{SPE}
Youngmoon Jung, Younggwan Kim, Hyungjun Lim, Yeunju Choi, and Hoirin Kim,
\newblock ``Spatial pyramid encoding with convex length normalization for
  text-independent speaker verification,''
\newblock in {\em Interspeech}, 2019, pp. 4030--4034.

\bibitem{richey2018voices}
Colleen Richey, Maria~A Barrios, Zeb Armstrong, Chris Bartels, Horacio Franco,
  Martin Graciarena, Aaron Lawson, Mahesh~Kumar Nandwana, Allen Stauffer,
  Julien van Hout, et~al.,
\newblock ``Voices obscured in complex environmental settings (voices)
  corpus,''
\newblock in {\em Interspeech}, 2018.

\bibitem{paszke2019pytorch}
Adam Paszke, Sam Gross, Francisco Massa, Adam Lerer, James Bradbury, Gregory
  Chanan, Trevor Killeen, Zeming Lin, Natalia Gimelshein, Luca Antiga, et~al.,
\newblock ``Pytorch: An imperative style, high-performance deep learning
  library,''
\newblock in {\em NeurIPS}, 2019, pp. 8024--8035.

\bibitem{nist2018}
{\em NIST 2018 Speaker Recognition Evaluation Plan}, 2018 (accessed 31 July
  2020),
\newblock
  \url{https://www.nist.gov/system/files/documents/2018/08/17/sre18_eval_plan_2018-05-31_v6.pdf},
  See Section 3.1.

\bibitem{chung2019voxsrc}
Joon~Son Chung, Arsha Nagrani, Ernesto Coto, Weidi Xie, Mitchell McLaren,
  Douglas~A Reynolds, and Andrew Zisserman,
\newblock ``{VoxSRC} 2019: The first {VoxCeleb} speaker recognition
  challenge,''
\newblock {\em arXiv preprint arXiv:1912.02522}, 2019.

\end{thebibliography}

\end{document}